\documentclass[a4paper,12pt]{article}
\usepackage{epsfig}

\newcommand{\be}{\begin{equation}}
\newcommand{\ee}{\end{equation}}
\newcommand{\ba}{\begin{eqnarray}}
\newcommand{\ea}{\end{eqnarray}}
\newcommand{\ban}{\begin{eqnarray*}}
\newcommand{\ean}{\end{eqnarray*}}

\newcommand{\demi}{\frac{1}{2}}

\begin{document}

\title{\Large\bf Does Quantum Mechanics imply influences\\acting backward in time\\
in impact series experiments?}

\author{{\bf Antoine Suarez}\thanks{suarez@leman.ch}\\ Center for Quantum
Philosophy\\ The Institute for Interdisciplinary Studies\\ P.O. Box
304, CH-8044 Zurich, Switzerland}


\maketitle

\vspace{2cm}
\begin{abstract}
A real two-particle experiment is proposed in which one of the
particles undergoes two successive impacts on beam-splitters. It is
shown that the standard quantum mechanical superposition principle
implies the possibility of influences acting backward in time
("retrocausation"), in striking contrast with the principle of
causality. It is argued that nonlocality and retrocausation are not
necessarily entangled.\\


\end{abstract}

\pagebreak

\section{Introduction}

Bell experiments with time-like separated impacts at the splitters
have already been done \cite{rar94} demonstrating the same
correlations as for space-like separated ones. Consider such an
experiment in which the measurement on particle $2$ lies time-like
separated after the measurement on particle $1$. It is clear that
at the time particle $1$ produces its outcome value, it cannot
account for values of particle $2$ because such values do not exist
at all, from any observer's point of view. In this case which
measurement is made first and which after dos not depend on the
inertial frame. Therefore, in agreement with the principle that the
effects cannot exist before the causes, it is reasonable to assume
that the correlations appear because particle 1 chooses its outcome
without being influenced by the choice particle 2 will make, and
particle 2 chooses its outcome taking account of the choice
particle 1 has made.\\

The impossibility of influences acting backward in time (the
causality principle) is basic to any causal model, independently of
one accepts or rejects the impossibility of superluminal influences
(relativistic causality). In particular, the causality principle
has been unified with the relativity of simultaneity in a
consistent way to account for the superluminal nonlocal influences,
and the consequent violation of relativistic causality, which
happen in Bell experiments with space-like separated measuring
devices. The resulting model is referred to as Relativistic
Nonlocality (RNL) or Multisimultaneity \cite{Sca1,Sua1}. Assuming
multisimultaneous causality, RNL is at odds with Lorentz-invariance
\cite{asvs972}. And even though RNL agrees with QM for all
experiments already done, both theories conflict in their
predictions regarding new proposed experiments with fast moving
polarizers.\\

The opposite view to the causal one is undoubtedly
"retrocausation", i.e., the position admiting that decisions at
present can influence the past. "Retrocausation" has been developed
as a consistent Lorentz-invariant interpretation of ordinary QM by
O. Costa de Beauregard \cite{co97}. The discussion about the
possibility of influences acting backwards in time has been
recently stimulated by H. Stapp \cite{hs97}. The ongoing
controversy \cite{wu97, dm97, jf98, vm98} is highlighting that we
have not yet found an specific experiment allowing us to decide
between the causal view and retrocausation, in a similar way as
Bell experiments allow us to decide between local realism and
superluminal nonlocality.\\

In this paper a possible real experiment is discussed in which
ordinary QM leads to predictions which imply influences backward
into a timelike separated past, and therefore may contribute to
clarify whether nature behaves retrocausal or not.

\section{The experiment}

Consider the setup sketched in Fig.1. Photon pairs are emitted
through down-conversion from a source $S$. Photon 1 enters the left
hand side interferometer and impacts on beam-splitter BS$_{11}$
before being detected in either D$_{1}(+)$ or D$_{1}(-)$, while
photon 2 enters the 2-interferometer series on the right hand side
impacting successively on BS$_{21}$ and BS$_{22}$ before being
detected in either D$_{2}(+)$ or D$_{2}(-)$. Each interferometer
consists in a long arm of length $L$, and a short one of length
$l$. We assume as usual the path difference set to a value which
largely exceeds the coherence length of the photon pair light, but
which is still smaller than the coherence length of the pump laser
light.\\

\begin{figure}[t]
\centering\epsfig{figure=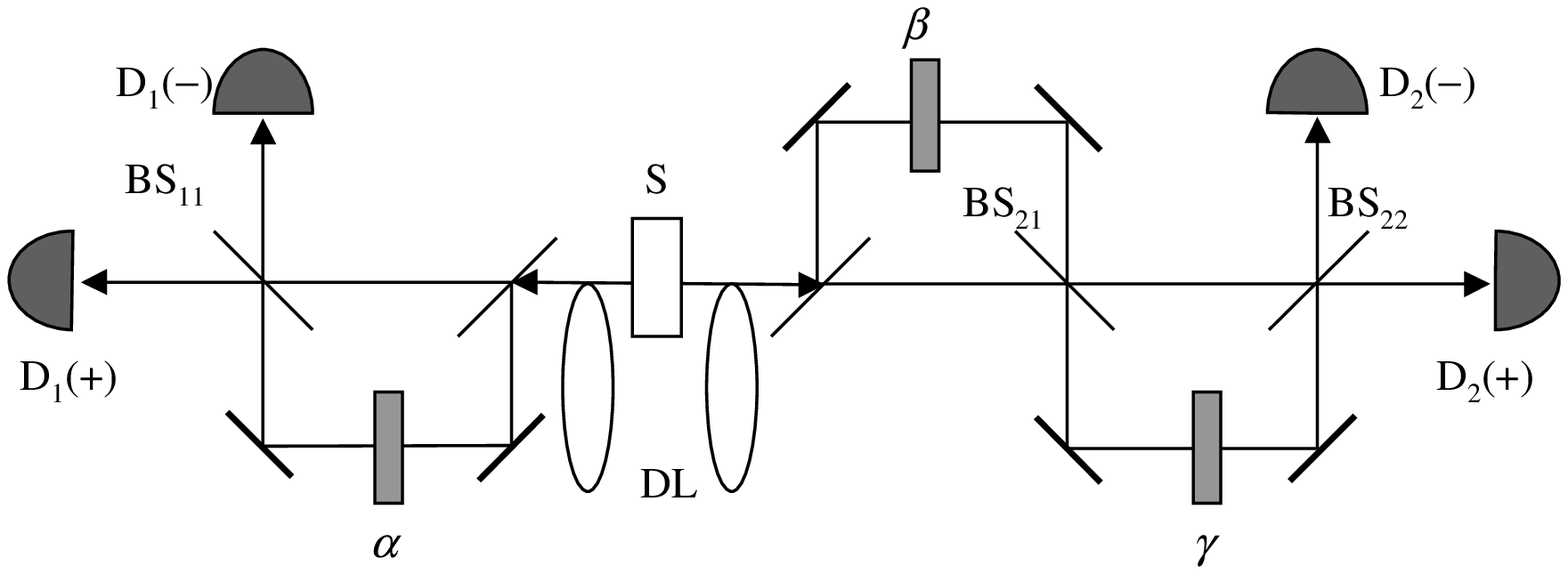,width=120mm}
{\small{\it{\caption{Impact series experiment with photon pairs:
photon 2 impacts successively on beam-splitter BS$_{21}$ and
BS$_{22}$. See text for detailed description.}}}}
\label{fig:BIPfig1}
\end{figure}

For a pair of photons, eight possible path pairs lead to detection.
We label them as follows: $(l,ll)$; $(L,ll)$; $(l,Ll)$ and so on;
where, e.g., $(l,Ll)$ indicates the path pair in wich photon 1 has
taken the short arm, and photon 2 has taken first the long arm,
then the short one.\\

Ordinary QM assumes indistinguishability to be a sufficient
condition for observing quantum interferences and entanglement,
whereas RNL assumes this condition to be only a necessary one. In
any case, as a first step we must distribute all possible paths in
mutually distinguishable subensembles. The following table gives
the four mutually distinguishable subensembles of the ensemble of
all possible path pairs.\\

\be
\begin{array}{lll}
(l,LL)&:&2L-l\\
(L,LL)\,,\,(l,Ll)\,,\,(l,lL)\,&:&L\\
(l,ll)\,,\,(L,Ll)\,,\,(L,lL)\,&:&l\\
(L,ll)&:&2l-L
\end{array}
\label{eq:paths}
\ee

where the right-hand side of the table indicates the path
difference between the single paths of each photon characterising
each subensemble of path pairs. From now on, unless stated
otherwise, we consider only those events that are characterized by
path difference $L$, i.e., $(L,LL)\,,\,(l,Ll)\,,\,(l,lL)$.
Experimentally, this is done as usual by appropriate coincidence
electronics \cite{tbg97}.\\

By means of delay lines DL the impacts on BS$_{11}$ are set
time-like separated from the impacts on BS$_{21}$ and BS$_{22}$. We
are interested in two different time orderings:

\begin{enumerate}
\item{The impact on BS$_{22}$ happens before the impact on
BS$_{11}$.}
\item{The impact on BS$_{11}$ happens before the impact on
BS$_{21}$.}
\end{enumerate}

\section{The QM view}

For reasons that will become clear in the next section we are not
interested in the joint probabilities but in the single
probabilities at each side of the setup, i.e., the probability of
getting a count in detector D$_{2}(\sigma)$ independently of where
photon 1 is detected, which we denote $P^{QM}_{\pm\sigma}(L)$, and
the probability of getting a count in detector D$_{1}(\sigma)$
independently of where photon 2 is detected, which we denote
$P^{QM}_{\sigma\pm}(L)$, where the $L$ in the parenthesis refers to
the corresponding path difference.\\

The single probabilities are related to the conventional joint ones
as follows:

\be
\begin{array}{lll}
P^{QM}_{\pm+}(L)\,\equiv\,P^{QM}_{++}(L)+P^{QM}_{-+}(L)\\
P^{QM}_{\pm-}(L)\,\equiv\,P^{QM}_{+-}(L)+P^{QM}_{--}(L)&
\label{eq:SPJP1}
\end{array}
\ee

and

\be
\begin{array}{lll}
P^{QM}_{+\pm}(L)\,\equiv\,P^{QM}_{++}(L)+P^{QM}_{+-}(L)\\
P^{QM}_{-\pm}(L)\,\equiv\,P^{QM}_{-+}(L)+P^{QM}_{--}(L)&
\label{eq:SPJP2}
\end{array}
\ee

Quantum mechanics is not time-ordering sensitive, and the
superposition principle states for any possible time ordering:

\ba
P^{QM}_{\sigma\omega}(L)&=&
\left|A_{\sigma\omega}(L,LL)+A_{\sigma\omega}(l,Ll)+
A_{\sigma\omega}(l,lL)\right|^2
\label{eq:JPQML}
\ea

where $A_{\sigma\omega}(path)$, ($\sigma,\omega\in\{+,-\}$), denote
the probability amplitudes for the path pair specified in the
parenthesis and the outcome specified in the subscript.
Substituting the amplitudes given in (\ref{Al,Ll}), (\ref{Al,lL})
and (\ref{AL,LL}) of the Appendix into Eq. (\ref{eq:JPQML}) and
adding according to (\ref{eq:SPJP1}) leads to the corresponding
single probabilities for the detections at side 2 (right-hand side)
of the setup:

\be
\begin{array}{lll}
P^{QM}_{\pm+}(L)&=&\demi+\frac{1}{3}\cos(\beta-\gamma)\\
P^{QM}_{\pm-}(L)&=&\demi-\frac{1}{3}\cos(\beta-\gamma)
\label{eq:SPQML1}
\end{array}
\ee

Adding according to (\ref{eq:SPJP2}) leads to the corresponding
single probabilities for the detections at side 1 (left-hand side)
of the setup:

\be
\begin{array}{lll}
P^{QM}_{+\pm}(L)&=&\demi-\frac{1}{3}\cos(\alpha+\beta)\\
P^{QM}_{-\pm}(L)&=&\demi+\frac{1}{3}\cos(\alpha+\beta)
\label{eq:SPQML2}
\end{array}\,
\ee

\smallskip

\section{The causal view}

According to the causal view, in experiments working with time
orderings 1 or 2 (see Section 2) the photon impacting before must
behave exclusively taking account of the local parameters, i.e., it
cannot become influenced by the choices of the parameters the other
photon meets at the other arm of the setup. This means for instance
in the experiment described in \cite{rar94} that the photon
impacting before produce single counts equally distributed, in
agreement with the predictions of QM and the observed results.\\

For reasons given in \cite{Sca1,Sua1} we consider in the following
that the outcome values for detections after beam-splitter
BS$_{ik}$ are determined at the time of arrival at this
beam-splitters, and not at the detectors watching the output ports
of BS$_{ik}$.\\

Consider now an experiment with time ordering 2. We accept that
whether photon 2 arriving at BS$_{21}$ or BS$_{22}$ undergoes a
transmission or a reflection may depend on which choice photon 1
did in BS$_{11}$, and hence on which D$_{1}(\sigma)$ it has been
detected. However to admit that the transmitted output port in
BS$_{21}$ corresponds necessarily to a short or a long arm in the
interferometer means to accept retrocausation, for the physicist is
always free to decide to shorten or lengthen the arm once photon 2
has made its choice. Accordingly we state the following
condition:\\

{\em Causality condition}: The path length traveled by the photon
impacting later does not depend on the outcome value produced by
the photon impacting first.\\

This condition implies that the distribution of the counts in the
single detectors produced by the photon impacting first, say photon
1, does not depend on the subensemble of path pairs in table
(\ref{eq:paths}) to which the event will belong once the detection
of photon 2 has occurred. In other words, even if the measurement
selects only those counts in the detectors D$_{1}(\sigma)$ yielding
path difference $L$ through coincidence with the counts in the
detectors D$_{2}(\omega)$, the measured distribution of the
outcomes in D$_{1}(\sigma)$ is the same as if it had been possible
to perform the experiment nonselectively with only the three paths
belonging to the subensemble $L$.\\

Taking account of this conclusion any causal model accepting the
available observations on first order interferences leads to the
following predictions:\\

{\em Time ordering 1}: After photon 2 impacts on BS$_{22}$ no
ulterior detection makes it possible to distinguish between the
paths (lL) and (Ll), but it is still possible to know whether
photon 2 traveled path (LL) by detecting particle 1 before it
impacts on BS$_{11}$. Therefore, if photon 2 behaves taking account
only of local information, paths (lL) and (Ll) lead to first order
interferences, and path (LL) does not interfere at all. The usual
application of the sum-of-probability-amplitudes and the
sum-of-probabilities leads to the relation:

\ba
P^{C}_{\pm\sigma}=
\left|A_{\sigma}(LL)\right|^2+\left|A_{\sigma}(Ll)+
A_{\sigma}(lL)\right|^2
\label{eq:proba1}
\ea

where $P^{C}_{\pm\sigma}$ denotes the single probability of getting
a count in detector D$_{2}(\sigma)$ predicted by the causal view,
and $A_{\sigma}(path)$ the amplitude associated with this detection
for the single path of photon 2 specified in the parenthesis.
Substituting according to (\ref{eq:ALl}), (\ref{eq:AlL}), and
(\ref{eq:ALL}) in the Appendix one gets the following single
probabilities for each detector D$_{2}(\sigma)$:

\be
\begin{array}{lll}
P^{C}_{\pm+}&=&\demi+\frac{1}{3}\cos(\beta-\gamma)\\
P^{C}_{\pm-}&=&\demi-\frac{1}{3}\cos(\beta-\gamma)
\label{eq:PC1}
\end{array}
\ee

i.e. one gets the same probabilities as those predicted by QM in
(\ref{eq:SPQML1}).\\

{\em Time ordering 2}: After the impact of photon 1 on BS$_{11}$ it
is still possible to know whether it traveled path $L$ or $l$ by
detecting photon 2 before it impacts on BS$_{21}$. Therefore photon
1 has to distribute its choices following the sum-of-probabilites
rule and one is led to the following probabilities:

\be
\begin{array}{lll}
P^{C}_{+\pm}&=&\demi\\
P^{C}_{-\pm}&=&\demi
\label{eq:PC2}
\end{array}
\ee

which clearly contradicts the QM predictions in
(\ref{eq:SPQML2}).\\

We would like to stress that the preceding result holds for any
theory accepting the causality principle, i.e., the impossibility
of influencing backward a timelike separated past. Obviously, one
would like to know also which probabilities predicts the causal
view for the single detections of photon 1 in time ordering 1, and
of photon 2 in time ordering 2. Notice that first of all this point
does not matter at all for our argument, and secondly these
probabilities will depend on the particular causal model under
consideration. As regards RNL or Mulsimultaneity \cite{Sca1, Sua1}
we give an answer in Section 6 below.

\section{Conflict between QM and Causality}

As far as we know this is the first time in a two-particle
experiment QM predicts single probabilities (\ref{eq:SPQML2}) for
one of the particles which depend on parameters the other particle
meets on the other side of the setup. The effect of retrocausation
violating the principle of causality is plain because it occurs
backward in time between timelike separated events.\\

Could such a retrocausation effect be used to built a time machine?
Consider the single probabilities for the subensemble with path
difference $l$ in Table (\ref{eq:paths}). The superposition
principle of QM states:

\ba
P^{QM}_{\sigma\omega}(l)&=&
\left|A_{\sigma\omega}(l,ll)+A_{\sigma\omega}(L,Ll)+
A_{\sigma\omega}(L,lL)\right|^2
\label{eq:proba3}
\ea

Substituting the amplitudes of (\ref{Al,ll}), (\ref{AL,lL}) and
(\ref{AL,Ll}) in the Appendix into Eq. (\ref{eq:proba3}) one gets:

\be
\begin{array}{lll}
P^{QM}_{+\pm}(L)&=&\demi+\frac{1}{3}\cos(\alpha+\beta)\\
P^{QM}_{-\pm}(L)&=&\demi-\frac{1}{3}\cos(\alpha+\beta)
\label{eq:SPQMl2}
\end{array}\,
\ee

Eq. (\ref{eq:SPQML2}) and (\ref{eq:SPQMl2}) together show that an
observer watching only the detectors D$_{1}$ cannot become aware in
the present of actions performed in the future of his light cone.
However, according to QM the coincidences measurement should
demonstrate such influences acting really backward in time. The
similarity with the superluminal nonlocality implied by ordinary QM
is impressive: in this case the coincidences measurement
demonstrates real faster-than-light influences, even though these
influences cannot be used for superluminal telegraphing.

\section{The RNL or mulsimultaneous causal view}

According to this theory \cite{Sua1} probabilities for counts in
single detectors must depend exclusively on local information,
i.e., they are the same for a {\em before} and a {\em non-before}
impact. Since in the proposed experiment the
sum-of-probability-amplitudes rule violates this principle, the
probabilites have to be calculate applying sum-of-probabilities. In
other words the violation of causality works in RNL in the same way
as the violation of indistinguishability in QM. Accordingly
(\ref{eq:PC1}) and (\ref{eq:PC2}) hold for the two considered time
orderings, and also for any other time ordering in experiments with
spacelike separated impacts: the presence of paths (lL) and (Ll)
leading to first order interferences excludes in this case the
second order ones.\\

Notice that RNL, though causal, is a specific nonlocal theory. That
it conflicts with QM suggests that the issues of superluminal
nonlocality and of retrocausation are not really entangled, and
should be conceptually distinguished: Nothing speaks in principle
against the possibility that Nature uses faster-than-light
influences but avoids backward-in-time ones.

\section{Real experiment}

A real experiment can be carried out arranging the setup used in
\cite{rar94} in order that the photon traveling the long fiber of
4.3 km impacts on a second beam-splitter before it is getting
detected. For the values:

\ba
\alpha+\beta=n\,\pi\,
\label{eq:re1}
\ea

with $n$ integer, the equations (\ref{eq:SPQML2}) and
(\ref{eq:PC1}) yield the predictions:

\ba
E^{QM}=|P^{QM}_{+\pm}(L)-P^{QM}_{-\pm}(L)|=\frac{2}{3}\nonumber\\
E^{C}=|P^{C}_{+\pm}(L)-P^{C}_{-\pm}(L)|=0
\label{eq:re2}
\ea

Hence, for settings according to (\ref{eq:re1}) the experiment
represented in Fig. 1 allow us to decide between quantum mechanics
and the causal view through determining the experimental quantity:

\ba
E=\frac{R_{++}+R_{+-}-R_{-+}+R_{--}}{R_{++}+R_{+-}+R_{-+}+R_{--}},
\label{eq:Ere}
\ea

where $R_{\sigma\omega}$ are the four measured coincidence counts
in the detectors.\\

\section{Conclusion}

We have shown that in the proposed impact series experiment
ordinary QM leads to influences backward in time, even if these
influences cannot be used to build a time machine. If the
experiment upholds QM, Costa de Beauregard's and Stapp's views
would appear to be the correct way of interpreting QM, quite in
agreement with Lorentz-invariance but in striking contradiction to
the causality principle. If the experiment upholds causality, then
Relativistic Non-Locality (RNL) or Multisimultaneity would receive
strong support. In RNL, indistinguishability is no more a
sufficient condition for entanglement, and both superluminal
influences as well as the impossibility of influences acting
backwards in time have the status of principles. Whatever the
answer may be, the experiment is capable of bearing a promising
controversy between QM and Causality, similar to the controversy
between QM and Local Realism.\\

\section*{Acknowledgements}

I would like to thank Valerio Scarani (EPFL, Lausanne) and Wolfgang
Tittel (University of Geneva) for numerous suggestions, and Olivier
Costa de Beauregard (L. de Broglie Foundation, Paris) for
stimulating discussions on retrocausation. It is a pleasure to
acknowledge also discussions regarding experimental realizations
with Nicolas Gisin and Hugo Zbinden (University of Geneva), and
with John Rarity and Paul Tapster (DRA, Malvern), and support by
the L\'eman and Odier Foundations.

\section*{Appendix}

In the following are listed the probability amplitudes of the path
pairs and the single paths we are interested in.

\subsection{Probability Amplitudes of the path pairs with
length difference $L$ in Table (\ref{eq:paths})}

We denote $A_{\sigma\omega}(\mbox{path})$ the probability amplitude
associated to detection of photon 1 in $D_{1}(\sigma)$ and of
photon 2 in $D_{2}(\omega)$, for the specified path. The
probability amplitudes for the path pairs of subensemble $L$ in
(\ref{eq:paths}) normalized to only these three path pairs are:

\ba
(l,Ll)&:&\left\{\begin{array}{lllll}
A_{++}(l,Ll)&=&-\,A_{--}(l,Ll)&=&-\frac{1}{\sqrt3}\,\frac{1}{2}\,e^{i\beta}\\
A_{+-}(l,Ll)&=&A_{-+}(l,Ll)&=&-i\,\frac{1}{\sqrt3}\,\frac{1}{2}\,e^{i\beta}
\label{Al,Ll}
\end{array}
\right.
\\
(l,lL)&:&\left\{\begin{array}{lllll}
A_{++}(l,lL)&=&A_{--}(l,lL)&=&-\frac{1}{\sqrt3}\,\frac{1}{2}\,e^{i\gamma}\\
A_{+-}(l,lL)&=&-\,A_{-+}(l,lL)&=&i\,\frac{1}{\sqrt3}\,\frac{1}{2}\,e^{i\gamma}
\label{Al,lL}
\end{array}
\right.
\\
(L,LL)&:&\left\{\begin{array}{lllll}
A_{++}(L,LL)&=&-\,A_{--}(L,LL)&=&\frac{1}{\sqrt3}\,\frac{1}{2}\,e^{i(\alpha+\beta+\gamma)}\\
A_{+-}(L,LL)&=&A_{-+}(L,LL)&=&-i\,\frac{1}{\sqrt3}\,\frac{1}{2}\,e^{i(\alpha+\beta+\gamma)}
\end{array}
\right.
\label{AL,LL}
\ea

\subsection{ Probability Amplitudes of the path pairs with
length difference $l$ in Table (\ref{eq:paths})}

The probability amplitudes for the path pairs of subensemble $l$ in
(\ref{eq:paths}) normalized to only these three path pairs are:

\ba
(l,ll)&:&\left\{\begin{array}{lllll}
A_{++}(l,ll)&=&-\,A_{--}(l,ll)&=&\frac{1}{\sqrt3}\,\frac{1}{2}\\
A_{+-}(l,ll)&=&A_{-+}(l,ll)&=&\frac{1}{\sqrt3}\,\frac{1}{2}\,i
\label{Al,ll}
\end{array}
\right.
\\
(L,lL)&:&\left\{\begin{array}{lllll}
A_{++}(L,lL)&=&-\,A_{--}(L,lL)&=&\frac{1}{\sqrt3}\,\frac{1}{2}\,e^{i(\alpha+\gamma)}\\
A_{+-}(L,lL)&=&A_{-+}(L,lL)&=&-\,\frac{1}{\sqrt3}\,\frac{1}{2}\,i\,e^{i(\alpha+\gamma)}
\label{AL,lL}
\end{array}
\right.
\\
(L,Ll)&:&\left\{\begin{array}{lllll}
A_{++}(L,Ll)&=&A_{--}(L,Ll)&=&\frac{1}{\sqrt3}\,\frac{1}{2}\,e^{i(\alpha+\beta)}\\
A_{+-}(L,Ll)&=&-\,A_{-+}(L,Ll)&=&\frac{1}{\sqrt3}\,\frac{1}{2}\,i\,e^{i(\alpha+\beta)}
\label{AL,Ll}
\end{array}
\right.
\ea

\subsection{Probability Amplitudes of the single paths $LL$, $Ll$, $lL$
traveled by photon 2 in the proposed experiment}

We denote $A_{\sigma}(\mbox{path})$ the probability amplitude
associated to detection of photon 2 in D$_{2}(\sigma)$, for the
specified path. The probability amplitudes for the paths $LL$,
$Ll$, $lL$ photon 2 travels in an experiment selecting the path
pairs with path difference $L$ in (\ref{eq:paths}), normalized as
if the experiment were performed with only these three paths are:

\ba
(Ll)&:&\left\{\begin{array}{lll}
A_{+}(Ll)&=&-\frac{1}{\sqrt3}\frac{1}{\sqrt2}\,e^{i\beta}\\
A_{-}(Ll)&=&-i\,\frac{1}{\sqrt3}\frac{1}{\sqrt2}\,e^{i\beta}
\label{eq:ALl}
\end{array}
\right.
\\
(lL)&:&\left\{\begin{array}{lll}
A_{+}(lL)&=&-\frac{1}{\sqrt3}\frac{1}{\sqrt2}\,e^{i\gamma}\\
A_{-}(lL)&=&i\,\frac{1}{\sqrt3}\frac{1}{\sqrt2}\,e^{i\gamma}
\label{eq:AlL}
\end{array}
\right.
\\
(LL)&:&\left\{\begin{array}{lll}
A_{+}(LL)&=&-\frac{1}{\sqrt3}\frac{1}{\sqrt2}\,e^{i(\beta+\gamma)}\\
A_{-}(LL)&=&i\,\frac{1}{\sqrt3}\frac{1}{\sqrt2}\,e^{i(\beta+\gamma)}
\label{eq:ALL}
\end{array}
\right.
\ea

\end{document}